\documentclass[sigchi-a,nonacm]{acmart}
\usepackage{booktabs} 

\usepackage{enumitem}
\usepackage{ragged2e}

\setcopyright{none}

\acmConference[CHI'19 Workshop on New Directions for the IoT: Automate, Share, Build, and Care]{}{May 2019}{Glasgow, UK}
\acmYear{2019}
\copyrightyear{2019}

\makeatletter
\def\@copyrightspace{\relax}
\renewcommand\@formatdoi[1]{\ignorespaces}
\makeatother

\begin{document}
\def\UrlFont{\small} 
\mathchardef\UrlBreakPenalty=10000

\title{Informing The Future of Data Protection in Smart Homes}

\author{Martin J Kraemer}
\affiliation{%
  \institution{Dept. of Computer Science\\University of Oxford}
  }
\email{martin.kraemer@cs.ox.ac.uk}

\author{William Seymour}
\affiliation{%
  \institution{Dept. of Computer Science\\University of Oxford}
  }
\email{william.seymour@cs.ox.ac.uk}

\author{Reuben Binns}
\affiliation{%
  \institution{Dept. of Computer Science\\University of Oxford}
  }
\email{reuben.binns@cs.ox.ac.uk}

\author{Max Van Kleek}
\affiliation{%
  \institution{Dept. of Computer Science\\University of Oxford}
  }
\email{max.van.kleek@cs.ox.ac.uk}

\author{Ivan Flechais}
\affiliation{%
  \institution{Dept. of Computer Science\\University of Oxford}
  }
\email{ivan.flechais@cs.ox.ac.uk}

\renewcommand{\shortauthors}{Kraemer et al.}

\begin{abstract}
Recent changes to data protection regulation, particularly in Europe, are changing the design landscape for smart devices, requiring new design techniques to ensure that devices are able to adequately protect users' data. A particularly interesting space in which to explore and address these challenges is the smart home, which presents a multitude of difficult social and technical problems in an intimate and highly private context.  This position paper outlines the motivation and research approach of a new project aiming to inform the future of data protection by design and by default in smart homes through a combination of ethnography and speculative design. 
\end{abstract}


\maketitle

\section{Introduction}
The General Data Protection Regulation elevated design guidelines referred to as Data Protection by Design and by Default to legal requirements within the EU. Manufacturers and designers are now required to consider data protection during the design process, making sure they comply with fundamental principles and requirements. Yet, it is non-obvious how such value-driven and abstract goals are to be translated into actual design requirements, let alone what designs that achieve such goals would look like, e.g.~\cite{Ceross2018,Wong2019}.

Meanwhile, bad practices by large companies continue to come to light through whistle blowers and data breaches, highlighting the power imbalance between companies and their users, with the resulting public outcry underlining the need for better privacy protection. If this imbalance is to be addressed, and if companies are to comply with legal requirements, then more informed ways of approaching data protection are required~\cite{Kraemer2018,Wong2019}. Of all the contexts that data protection by design applies to, the home is a particularly interesting place to research these issues because of its fundamental position in social life and the fact that smart home devices are becoming increasingly ubiquitous and unobtrusive~\cite{Kraemer2018,Wilson2015}.

\begin{sidebar}
\section{What does DPbD/D mean for businesses?}
Excerpt of guidelines from the UK Information Commissioner's Office (ICO)
\small
\begin{itemize}[leftmargin=*]
    \item Consider data protection part of design and implementation of systems, services, products and business practices.
    \item Make data protection a core functionality of data processing
    \item Anticipate and prevent privacy-invasive events before they occur
    \item Only process and use personal data that is needed for the organisation's purposes(s)
    \item Ensure that personal data is automatically protected in any IT system, service, product, and/or business practice
    \item Adopt a `plain language' policy for any public documents
    \item Provide individuals with tools so they can determine how their personal data is being used
    \item Offer strong privacy defaults, user-friendly options and controls, and respect user preferences.
\end{itemize}
\end{sidebar}

Research on smart device usage in the home has highlighted how dichotomies between users and administrators can also lead to power imbalances~\cite{Geeng2019}. Configuration and maintenance of smart devices is typically carried out by a specific householder while the use of these devices might be shared with many, leading to conflicts and tensions between users~\cite{Geeng2019,Nthala2018}. To avoid such situations, promoting inclusivity and collaboration between inhabitants through product design is paramount. While novel technology is often reported to be the cause of these issues, it also offers the key to addressing and preventing them.

In this position paper we present an outline of our ongoing research project which seeks to inform the future of Data Protection by Design and by Default (DPbD/D) in smart homes. We plan to achieve this by informing speculative design through ethnography and discussions with industry experts at partner organisations to map out the design landscape offered by DpbD/D. Specifically, we aim to:

\begin{enumerate}
    \item better understand the social embeddedness of technology use, individual and communal privacy practices
    \item develop and prototype reusable artefacts for DpbD/D
    \item evaluate and extend our findings in collaboration with industry and regulatory stakeholders to create applicable and actionable outputs
\end{enumerate}

\section{Background \& Prior Work}

The GDPR's principle of data protection by design and by default (see sidebar) is based on previous notions of privacy by design~\cite{Cavoukian2010}, with the aim of designing for higher-level values such as autonomy and privacy~\cite{Wong2019,Ceross2018}. A recent literature review of HCI design contributions for Privacy by Design highlights how many existing contributions used design to either solve a problem or to support and inform privacy decision making~\cite{Wong2019}.

We agree that it is valuable for researchers to ``explore people and situations and to critique, speculate, or present critical alternatives'', especially where this is used to unravel ``entangled relationships among the social, technical, and legal''~\cite{Wong2019}. This resonates well with others who reviewed the emerging field of privacy engineering and identified the need for more contextual research to fill in the gaps between principle driven privacy regulation, a user-centred socio-technical perspective, and feature driven or problem-solving engineering practices~\cite{Ceross2018}. We address the need for this research through our focus on \textit{individual and communal privacy} for which, to the best of our knowledge, no theory of privacy exists~\cite{Wong2019,Kraemer2018}.

This need is illustrated by Lau et al. in their paper on smart voice assistants, where they argue the need for better privacy choice controls, a form of privacy enhancing technology (PET)~\cite{Lau2018}. PETs in user interfaces of devices allow individuals to state preferences for data collection, processing, and dissemination practices~\cite{Schraefel2017}. Despite these efforts, reports of questionable company practices and data leaks continue to cause power imbalances between users and manufacturers~\cite{Zheng2018a,Crabtree2016b}; on learning about data practices and repercussions of data breaches, experts and non-experts alike express feelings of ``bewilderment, resistance, and sometimes resignation''~\cite[p. 3]{Nissenbaum2009}. Prior work has identified related challenges for the individual user, e.g. abstract and a priori decision making~\cite{acquisti2005privacy}, grasping the meaning of these settings~\cite{Schraefel2017}, and challenges of appropriating products to fit their needs~\cite{Lau2018}. The complexity of navigating these situations is exaggerated through communal use, embedded in existing social order and dynamics, e.g.~\cite{Zeng2017,Crabtree2015a,Crabtree2012}, and so a related power imbalance can be found within communities such as households sharing internet-connected devices~\cite{Zeng2017,Geeng2019}.

If Data Protection by Design and by Default is to become a tangible goal in software development and its benefits reality for users, then a holistic approach to understanding user needs and practices, producing artefacts for software and product design, and evaluating these with industrial and regulatory stakeholders are all required. Our prior research equips us with the required knowledge and skills to tackle this complex challenge.

\section{Research Approach}
Our previous research into transparency tools for smartphone apps \cite{van2018x}, smart home devices (under submission), as well as other prior work \cite{shklovski2014leakiness} has explored the mental models and coping strategies that are used to navigate the complex information flows generated by smart devices in the home.


We have also explored the space of home technology use and support in our prior research from a security perspective~\cite{Nthala2018}. Other prior work has also reviewed existing privacy literature to devise a road map for privacy research in the home~\cite{Kraemer2018}. From this, we have taken the initial step of exploring smart device usage through a series of interviews focusing on questions of procurement, daily use, and problem solving with internet-connected devices. We are currently working on using these insights to improve contextual design approaches and techniques with regards to usable security and privacy (under submission).


With the above in mind, the project aims to further explore the unique privacy needs arising in smart homes, as well as the design techniques required to meet those needs. The three main stages of the project are as follows:

\begin{sidebar}
	\vfill
	
	\section{Ethnographic Study Design}
	\vspace{0.5cm}
	
	\centering
	\includegraphics[width=0.45\columnwidth]{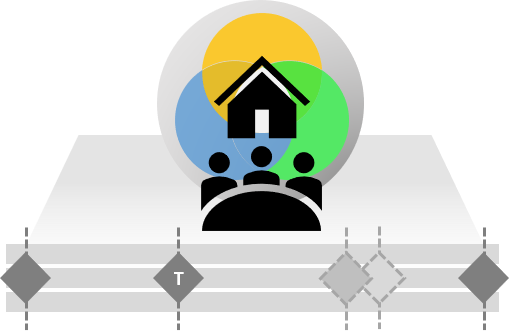}
	\begin{center}
		Ethnographic interview ($\diamond$) study with ethnomethodology as analytic lens \\ (n = 5-8 households)
	\end{center}

	\par\vspace{1ex}
	\begin{flushleft}
		Our visits will focus on
		\begin{enumerate}[label=(\Alph*)]
			\item existing practices with internet-connected, smart technology
			\item process of negotiating device placement, configuration, and potentially usage---after households choose from a pool of \textit{'invasive'} devices, removing the economic entry barrier (see (T))
			\item[(C/D)] routine use of devices and problem solving in relation to privacy as such practices evolve over time
		\end{enumerate}
	\end{flushleft}
	\vfill
\end{sidebar}
\begin{enumerate}
    \item An ethnographic longitudinal study to explore and understand individual and communal digital privacy practices pertaining to smart home devices and their privacy choice controls
    \item The prototyping of new tools, interfaces, and approaches to smart home privacy, informed by the longitudinal study, including co-design sessions with smart home users
    \item Discussions with smart home product designers, product teams, and compliance officers to understand how these alternative design approaches might be integrated into their DPbD/D processes
\end{enumerate}

\subsection{Individual and Communal Digital Privacy Practices}
To disentangle the problem space of individual and communal digital privacy practices of smart home device use, we propose an ethnographic approach with ethnomethodology as analytic lens~\cite{Crabtree2012a,Crabtree2017a}. We choose ethnomethodology as analytic lens because of its focus on how members of the household accomplish their goals and how they consider others in so doing, i.e. how they make their own actions accountable; and because of its value to system design and software development~\cite{Crabtree2012a}.

We will spend 6 months working with participating households, learning about their experience with existing and new smart technology (see sidebar). For this purpose, we offer households a number of smart devices from a pool of selected hardware---the mundanity that we are interested in is the use of internet-connected technologies embedded in social life. Ethnography with ethnomethodology as analytic lens allows us to observe the social embeddedness~\cite{Crabtree2015a,Goulden2018} of technology use---``Humans are here to stay. Technologies come and go'' -- Tom Rodden, \cite[p. 14]{Crabtree2012a}.

\subsection{Prototyping and Evaluating Design Artefacts}
The second part of the project involves rapidly prototyping devices and interfaces that embody the implementation of data protection by design and by default in smart home devices, as informed by the results of the intial phase of the project. This space will be explored through iterative conceptual ideation, prototyping, development, followed by lightweight evaluation of a range of speculative prototypes, encompassing a range of different interfaces, features and UI design patterns. A small number of the initial outlines will be developed into functional prototypes.

The aim of these prototypes is to understand how alternative approaches to DPbD/D might better serve the needs identified of device users. We will also be exploring how alternative forms of transparency can help users develop useful and functional mental models of the internal logic and data processing practices of smart devices. By creating prototypes that back up transparency with well reasoned follow on actions, we aim to support users in developing and realising their personal privacy preferences, including through better use of their legal rights as data subjects. 

\subsection{Integrating DPbD/D into Product Design}
To identify and develop a range of DPbD patterns from an industry perspective, the final stage of the project will use focus groups targeted at developers, designers, and data protection compliance officers involved in the design of smart home devices and services to understand how the insights and artefacts that have been produced through stages one and two can be incorporated into the product lifecycle. Participants for the focus groups will be drawn from our industrial partners.

\section{Outlook}
\begin{sidebar}
	\section{Acknowledgements}
	This research is part of our initiative `Informing the Future of Data Protection by Design and by Default in Smart Homes' at the University of Oxford.
	Martin Kraemer and William Seymour are supported by the UK Engineering and Physical Sciences research council (EPSRC) through grant number P00881X/1.\\\vspace{1em}

	Our project page:\\
	\url{https://www.cs.ox.ac.uk/projects/fosh/}
\end{sidebar}
Effectively investigating the perceptions and behaviours that shape how people use IoT devices in the most private space in their lives is fraught with challenges. Through the combination of an HCI research approach with aspects of social science methodology and speculative design, we hope to explore the highly complex and idiosyncratic issues posed by the use of smart devices in the home. At the workshop, we hope to discuss the challenges around inclusively and effectively involving all users in the home with DPbD/D, sharing our insights with others and soliciting feedback on our own methods.

\bibliography{refs.bib}


\begin{thebibliography}{00}


\ifx \showCODEN    \undefined \def \showCODEN     #1{\unskip}     \fi
\ifx \showDOI      \undefined \def \showDOI       #1{#1}\fi
\ifx \showISBNx    \undefined \def \showISBNx     #1{\unskip}     \fi
\ifx \showISBNxiii \undefined \def \showISBNxiii  #1{\unskip}     \fi
\ifx \showISSN     \undefined \def \showISSN      #1{\unskip}     \fi
\ifx \showLCCN     \undefined \def \showLCCN      #1{\unskip}     \fi
\ifx \shownote     \undefined \def \shownote      #1{#1}          \fi
\ifx \showarticletitle \undefined \def \showarticletitle #1{#1}   \fi
\ifx \showURL      \undefined \def \showURL       {\relax}        \fi
\providecommand\bibfield[2]{#2}
\providecommand\bibinfo[2]{#2}
\providecommand\natexlab[1]{#1}
\providecommand\showeprint[2][]{arXiv:#2}

\bibitem[\protect\citeauthoryear{Acquisti and Grossklags}{Acquisti and
  Grossklags}{2005}]%
        {acquisti2005privacy}
\bibfield{author}{\bibinfo{person}{Alessandro Acquisti} {and}
  \bibinfo{person}{Jens Grossklags}.} \bibinfo{year}{2005}\natexlab{}.
\newblock \showarticletitle{Privacy and rationality in individual decision
  making}.
\newblock \bibinfo{journal}{{\em IEEE security \& privacy\/}}
  \bibinfo{volume}{3}, \bibinfo{number}{1} (\bibinfo{year}{2005}),
  \bibinfo{pages}{26--33}.
\newblock


\bibitem[\protect\citeauthoryear{Cavoukian}{Cavoukian}{2010}]%
        {Cavoukian2010}
\bibfield{author}{\bibinfo{person}{Ann Cavoukian}.}
  \bibinfo{year}{2010}\natexlab{}.
\newblock \bibinfo{booktitle}{{\em {The 7 Foundational Principles
  Implementation and Mapping of Fair Information Practices}}}.
\newblock \bibinfo{type}{{T}echnical {R}eport}.
  \bibinfo{institution}{Information and Privacy Commissioner of Ontario}.
  \bibinfo{pages}{10} pages.
\newblock
\showURL{%
\url{www.ipc.on.ca/images/Resources/gps.pdf}}


\bibitem[\protect\citeauthoryear{Ceross and Simpson}{Ceross and
  Simpson}{2018}]%
        {Ceross2018}
\bibfield{author}{\bibinfo{person}{Aaron Ceross} {and} \bibinfo{person}{Andrew
  Simpson}.} \bibinfo{year}{2018}\natexlab{}.
\newblock \showarticletitle{{Rethinking the Proposition of Privacy
  Engineering}}. In \bibinfo{booktitle}{{\em Proceedings of the 2018 New
  Security Paradigms Workshop}}. \bibinfo{publisher}{ACM},
  \bibinfo{pages}{forthcoming}.
\newblock
\showISBNx{9781450365970}


\bibitem[\protect\citeauthoryear{Crabtree and Mortier}{Crabtree and
  Mortier}{2016}]%
        {Crabtree2016b}
\bibfield{author}{\bibinfo{person}{Andy Crabtree} {and}
  \bibinfo{person}{Richard Mortier}.} \bibinfo{year}{2016}\natexlab{}.
\newblock \showarticletitle{{Personal Data, Privacy and the Internet of Things:
  The Shifting Locus of Agency and Control}}.
\newblock \bibinfo{journal}{{\em SSRN\/}} (\bibinfo{year}{2016}),
  \bibinfo{pages}{1--20}.
\newblock
\showURL{%
\url{https://ssrn.com/abstract=2874312}}


\bibitem[\protect\citeauthoryear{Crabtree, Mortier, Rodden, and
  Tolmie}{Crabtree et~al\mbox{.}}{2012a}]%
        {Crabtree2012}
\bibfield{author}{\bibinfo{person}{Andy Crabtree}, \bibinfo{person}{Richard
  Mortier}, \bibinfo{person}{Tom Rodden}, {and} \bibinfo{person}{Peter
  Tolmie}.} \bibinfo{year}{2012}\natexlab{a}.
\newblock \showarticletitle{{Unremarkable networking: the home network as a
  part of everyday life}}, In \bibinfo{booktitle}{Proceedings of the Designing
  Interactive Systems Conference}.
\newblock \bibinfo{journal}{{\em Proceedings of the Designing Interactive
  Systems Conference. ACM.\/}}, \bibinfo{pages}{554--563}.
\newblock


\bibitem[\protect\citeauthoryear{Crabtree, Rodden, Tolmie, Mortier, Lodge,
  Brundell, and Pantidi}{Crabtree et~al\mbox{.}}{2015}]%
        {Crabtree2015a}
\bibfield{author}{\bibinfo{person}{Andy Crabtree}, \bibinfo{person}{Tom
  Rodden}, \bibinfo{person}{Peter Tolmie}, \bibinfo{person}{Richard Mortier},
  \bibinfo{person}{Tom Lodge}, \bibinfo{person}{Pat Brundell}, {and}
  \bibinfo{person}{Nadia Pantidi}.} \bibinfo{year}{2015}\natexlab{}.
\newblock \showarticletitle{{House rules: the collaborative nature of policy in
  domestic networks}}.
\newblock \bibinfo{journal}{{\em Personal and Ubiquitous Computing\/}}
  \bibinfo{volume}{19}, \bibinfo{number}{1} (\bibinfo{year}{2015}),
  \bibinfo{pages}{203--215}.
\newblock


\bibitem[\protect\citeauthoryear{Crabtree, Rouncefield, and Tolmie}{Crabtree
  et~al\mbox{.}}{2012b}]%
        {Crabtree2012a}
\bibfield{author}{\bibinfo{person}{Andy Crabtree}, \bibinfo{person}{Mark
  Rouncefield}, {and} \bibinfo{person}{Peter Tolmie}.}
  \bibinfo{year}{2012}\natexlab{b}.
\newblock \bibinfo{booktitle}{{\em {Doing Design Ethnography}\/}
  (\bibinfo{edition}{1} ed.)}.
\newblock \bibinfo{publisher}{Springer-Verlag London}.
\newblock


\bibitem[\protect\citeauthoryear{Crabtree, Tolmie, and Knight}{Crabtree
  et~al\mbox{.}}{2017}]%
        {Crabtree2017a}
\bibfield{author}{\bibinfo{person}{Andy Crabtree}, \bibinfo{person}{Peter
  Tolmie}, {and} \bibinfo{person}{Will Knight}.}
  \bibinfo{year}{2017}\natexlab{}.
\newblock \showarticletitle{{Repacking 'Privacy' for a Networked World}}.
\newblock \bibinfo{journal}{{\em Computer Supported Cooperative Work: CSCW: An
  International Journal\/}} \bibinfo{volume}{26}, \bibinfo{number}{4-6}
  (\bibinfo{year}{2017}), \bibinfo{pages}{453--488}.
\newblock


\bibitem[\protect\citeauthoryear{Geeng and Roesner}{Geeng and Roesner}{2019}]%
        {Geeng2019}
\bibfield{author}{\bibinfo{person}{Christine Geeng} {and}
  \bibinfo{person}{Franziska Roesner}.} \bibinfo{year}{2019}\natexlab{}.
\newblock \showarticletitle{{Who's In Control?: Interactions In Multi-User
  Smart Homes}}. In \bibinfo{booktitle}{{\em Proceedings of the SIGCHI
  Conference on Human Factors in Computing Systems - CHI'19}}.
  \bibinfo{publisher}{ACM}, \bibinfo{pages}{forthcoming}.
\newblock


\bibitem[\protect\citeauthoryear{Goulden, Tolmie, Mortier, Lodge, Pietilainen,
  and Teixeira}{Goulden et~al\mbox{.}}{2018}]%
        {Goulden2018}
\bibfield{author}{\bibinfo{person}{Murray Goulden}, \bibinfo{person}{Peter
  Tolmie}, \bibinfo{person}{Richard Mortier}, \bibinfo{person}{Tom Lodge},
  \bibinfo{person}{Anna~Kaisa Pietilainen}, {and} \bibinfo{person}{Renata
  Teixeira}.} \bibinfo{year}{2018}\natexlab{}.
\newblock \showarticletitle{{Living with interpersonal data: Observability and
  accountability in the age of pervasive ICT}}.
\newblock \bibinfo{journal}{{\em New Media and Society\/}}
  \bibinfo{volume}{20}, \bibinfo{number}{4} (\bibinfo{year}{2018}),
  \bibinfo{pages}{1580--1599}.
\newblock


\bibitem[\protect\citeauthoryear{Kraemer and Flechais}{Kraemer and
  Flechais}{2018}]%
        {Kraemer2018}
\bibfield{author}{\bibinfo{person}{Martin~J Kraemer} {and}
  \bibinfo{person}{Ivan Flechais}.} \bibinfo{year}{2018}\natexlab{}.
\newblock \showarticletitle{{Researching Privacy in Smart Homes : A Roadmap of
  Future Directions and Research Methods}}.
\newblock \bibinfo{journal}{{\em IET Conference Proceedings\/}}
  (\bibinfo{year}{2018}), \bibinfo{pages}{1--10}.
\newblock


\bibitem[\protect\citeauthoryear{Lau, Zimmerman, and Schaub}{Lau
  et~al\mbox{.}}{2018}]%
        {Lau2018}
\bibfield{author}{\bibinfo{person}{Josephine Lau}, \bibinfo{person}{Benjamin
  Zimmerman}, {and} \bibinfo{person}{Florian Schaub}.}
  \bibinfo{year}{2018}\natexlab{}.
\newblock \showarticletitle{{Alexa, Are You Listening?: Privacy Perceptions,
  Concerns and Privacy-seeking Behaviors with Smart Speakers}}.
\newblock \bibinfo{journal}{{\em Proc. ACM Hum.-Comput. Interact.\/}}
  \bibinfo{volume}{2}, \bibinfo{number}{CSCW} (\bibinfo{date}{nov}
  \bibinfo{year}{2018}), \bibinfo{pages}{102:1----102:31}.
\newblock


\bibitem[\protect\citeauthoryear{{M.C. Schraefel}, {R. Gomer}, {A. Alan}, {E.
  Gerding}, and {C. Maple}}{{M.C. Schraefel} et~al\mbox{.}}{2017}]%
        {Schraefel2017}
\bibfield{author}{\bibinfo{person}{{M.C. Schraefel}}, \bibinfo{person}{{R.
  Gomer}}, \bibinfo{person}{{A. Alan}}, \bibinfo{person}{{E. Gerding}}, {and}
  \bibinfo{person}{{C. Maple}}.} \bibinfo{year}{2017}\natexlab{}.
\newblock \showarticletitle{{The Internet of Things: Interaction Challenges to
  Meaningful Consent at Scale}}.
\newblock \bibinfo{journal}{{\em Interactions\/}} \bibinfo{volume}{24},
  \bibinfo{number}{6} (\bibinfo{date}{Nov.} \bibinfo{year}{2017}),
  \bibinfo{pages}{27--33}.
\newblock


\bibitem[\protect\citeauthoryear{Nissenbaum}{Nissenbaum}{2009}]%
        {Nissenbaum2009}
\bibfield{author}{\bibinfo{person}{Helen Nissenbaum}.}
  \bibinfo{year}{2009}\natexlab{}.
\newblock \bibinfo{booktitle}{{\em {Privacy in context: Technology, policy, and
  the integrity of social life}}}.
\newblock \bibinfo{publisher}{Stanford University Press}.
\newblock


\bibitem[\protect\citeauthoryear{Nthala and Flechais}{Nthala and
  Flechais}{2018}]%
        {Nthala2018}
\bibfield{author}{\bibinfo{person}{Norbert Nthala} {and} \bibinfo{person}{Ivan
  Flechais}.} \bibinfo{year}{2018}\natexlab{}.
\newblock \showarticletitle{Informal Support Networks: an investigation into
  Home Data Security Practices}. In \bibinfo{booktitle}{{\em Fourteenth
  Symposium on Usable Privacy and Security ({SOUPS} 2018)}}.
  \bibinfo{publisher}{{USENIX} Association}, \bibinfo{address}{Baltimore, MD},
  \bibinfo{pages}{63--82}.
\newblock


\bibitem[\protect\citeauthoryear{Shklovski, Mainwaring, Sk{\'u}lad{\'o}ttir,
  and Borgthorsson}{Shklovski et~al\mbox{.}}{2014}]%
        {shklovski2014leakiness}
\bibfield{author}{\bibinfo{person}{Irina Shklovski}, \bibinfo{person}{Scott~D
  Mainwaring}, \bibinfo{person}{Halla~Hrund Sk{\'u}lad{\'o}ttir}, {and}
  \bibinfo{person}{H{\"o}skuldur Borgthorsson}.}
  \bibinfo{year}{2014}\natexlab{}.
\newblock \showarticletitle{Leakiness and creepiness in app space: Perceptions
  of privacy and mobile app use}. In \bibinfo{booktitle}{{\em Proceedings of
  the CHI Conference on Human Factors in Computing Systems}} {\em
  (\bibinfo{series}{CHI '14})}. ACM, \bibinfo{pages}{2347--2356}.
\newblock


\bibitem[\protect\citeauthoryear{Van~Kleek, Binns, Zhao, Slack, Lee, Ottewell,
  and Shadbolt}{Van~Kleek et~al\mbox{.}}{2018}]%
        {van2018x}
\bibfield{author}{\bibinfo{person}{Max Van~Kleek}, \bibinfo{person}{Reuben
  Binns}, \bibinfo{person}{Jun Zhao}, \bibinfo{person}{Adam Slack},
  \bibinfo{person}{Sauyon Lee}, \bibinfo{person}{Dean Ottewell}, {and}
  \bibinfo{person}{Nigel Shadbolt}.} \bibinfo{year}{2018}\natexlab{}.
\newblock \showarticletitle{X-Ray Refine: Supporting the Exploration and
  Refinement of Information Exposure Resulting from Smartphone Apps}. In
  \bibinfo{booktitle}{{\em Proceedings of the 2018 CHI Conference on Human
  Factors in Computing Systems}} {\em (\bibinfo{series}{CHI '18})}. ACM,
  \bibinfo{pages}{393}.
\newblock


\bibitem[\protect\citeauthoryear{Wilson, Hargreaves, and
  Hauxwell-Baldwin}{Wilson et~al\mbox{.}}{2015}]%
        {Wilson2015}
\bibfield{author}{\bibinfo{person}{Charlie Wilson}, \bibinfo{person}{Tom
  Hargreaves}, {and} \bibinfo{person}{Richard Hauxwell-Baldwin}.}
  \bibinfo{year}{2015}\natexlab{}.
\newblock \showarticletitle{{Smart homes and their users: a systematic analysis
  and key challenges}}.
\newblock \bibinfo{journal}{{\em Personal and Ubiquitous Computing\/}}
  \bibinfo{volume}{19}, \bibinfo{number}{2} (\bibinfo{year}{2015}),
  \bibinfo{pages}{463--476}.
\newblock


\bibitem[\protect\citeauthoryear{Wong and Mulligan}{Wong and Mulligan}{2019}]%
        {Wong2019}
\bibfield{author}{\bibinfo{person}{Richmond~Y Wong} {and}
  \bibinfo{person}{Deirdre~K Mulligan}.} \bibinfo{year}{2019}\natexlab{}.
\newblock \showarticletitle{{Bringing Design to the Privacy Table - Broadening
  "Design" in "Privacy by Design" Through the Lens of HCI}}. In
  \bibinfo{booktitle}{{\em Proceedings of the 2019 CHI Conference on Human
  Factors in Computing Systems - CHI '19}}. \bibinfo{publisher}{ACM},
  \bibinfo{pages}{forthcoming}.
\newblock
\showISBNx{9781450359702}
\showDOI{%
\url{https://doi.org/10.1145/3290605.3300492}}


\bibitem[\protect\citeauthoryear{Zeng, Mare, and Roesner}{Zeng
  et~al\mbox{.}}{2017}]%
        {Zeng2017}
\bibfield{author}{\bibinfo{person}{Eric Zeng}, \bibinfo{person}{Shrirang Mare},
  {and} \bibinfo{person}{Franziska Roesner}.} \bibinfo{year}{2017}\natexlab{}.
\newblock \showarticletitle{End User Security \& Privacy Concerns with Smart
  Homes}. In \bibinfo{booktitle}{{\em Proceedings of the Thirteenth USENIX
  Conference on Usable Privacy and Security}} {\em
  (\bibinfo{series}{SOUPS'17})}. \bibinfo{publisher}{USENIX Association},
  \bibinfo{address}{Berkeley, CA, USA}, \bibinfo{pages}{65--80}.
\newblock


\bibitem[\protect\citeauthoryear{Zheng, Apthorpe, Chetty, and Feamster}{Zheng
  et~al\mbox{.}}{2018}]%
        {Zheng2018a}
\bibfield{author}{\bibinfo{person}{Serena Zheng}, \bibinfo{person}{Noah
  Apthorpe}, \bibinfo{person}{Marshini Chetty}, {and} \bibinfo{person}{Nick
  Feamster}.} \bibinfo{year}{2018}\natexlab{}.
\newblock \showarticletitle{User Perceptions of Smart Home IoT Privacy}.
\newblock \bibinfo{journal}{{\em Proc. ACM Hum.-Comput. Interact.\/}}
  \bibinfo{volume}{2}, \bibinfo{number}{CSCW} (\bibinfo{date}{Nov.}
  \bibinfo{year}{2018}), \bibinfo{pages}{200:1--200:20}.
\newblock


\end{thebibliography}
\bibliographystyle{ACM-Reference-Format}

\end{document}